\documentclass[fleqn,10pt]{wlscirep}
\usepackage[utf8]{inputenc}
\usepackage[T1]{fontenc}
\usepackage{gensymb}
\usepackage{txfonts}
\usepackage{scalefnt}
\usepackage{color}

\title{X-ray free-electron lasing in a flying-focus undulator}

\author[1,*]{D. Ramsey}
\author[2]{B. Malaca}
\author[1]{T.T. Simpson}
\author[3]{M. Formanek}
\author[1]{L. S. Mack}
\author[2]{J. Vieira}
\author[1]{D.H. Froula}
\author[1]{J.P. Palastro}

\affil[1]{Laboratory for Laser Energetics, University of Rochester, Rochester, NY 14623-1299, USA}
\affil[2]{GoLP/Instituto de Plasmas e Fusão Nuclear, Instituto Superior Técnico, Universidade de Lisboa, Lisbon 1049-001, Portugal}
\affil[3]{ELI Beamlines Facility, The Extreme Light Infrastructure ERIC, Doln\'{i} B\v{r}e\v{z}any 252 41, Czech Republic}

\affil[*]{dram@lle.rochester.edu}

\newcommand{\unitvec}[1]{\ensuremath{\boldsymbol{\hat{\mathrm{#1}}}}}
\newcommand\Genesis{{\scalefont{0.85}GENESIS-1.3 }}
\newcommand\GenesisNoSpace{{\scalefont{0.85}GENESIS-1.3}}
\newcommand{\bvec}[1]{\ensuremath{\boldsymbol{\mathrm{#1}}}}

\begin{abstract}
Laser-driven free-electron lasers (LDFELs) replace magnetostatic undulators with the electromagnetic fields of a laser pulse. Because the undulator period is half the wavelength of the laser pulse, LDFELs can amplify x rays using lower electron energies and over shorter interaction lengths than a traditional free-electron laser. In LDFELs driven by conventional laser pulses, the undulator uniformity required for high gain necessitates large laser-pulse energies. Here, we show that a flying-focus pulse provides the undulator uniformity required to reach high gain with a substantially lower energy than a conventional pulse. The flying-focus pulse features an intensity peak that travels in the opposite direction of its phase fronts. This enables an LDFEL configuration where an electron beam collides head-on with the phase fronts and experiences a near-constant undulator strength as it co-propagates with the intensity peak. Three-dimensional simulations of this configuration demonstrate the generation of megawatts of coherent x-ray radiation with 20$\times$ less energy than a conventional laser pulse.
\end{abstract}

\begin{document}

\flushbottom
\maketitle

\thispagestyle{empty}

\section*{Introduction}

Sources of coherent x rays are vital to medical, engineering, and basic scientific research. Coherent x rays allow for phase-contrast and diffractive imaging of molecules, cells, high-energy-density materials, and structural defects\cite{wilkins1996phase,Pfeiffer2006,Barty2008,robinson2009coherent,miao2015beyond}; absorption spectroscopy and Thomson scattering to probe the structure and evolution of matter across phase changes\cite{glenzer2009x,yano2009x,ciricosta2012direct,vinko2012creation}; and the exploration and observation of quantum-electrodynamical processes, such as pair-production, photon--photon scattering, and vacuum birefringence\cite{ringwald2001pair,heinzl2006observation, RevModPhys.84.1177,gonoskov2022charged, fedotov2023advances, macleod2023strong}. The most-brilliant coherent sources reside at large-scale accelerator facilities, where high-energy electron beams fired into a magnetostatic undulator produce x rays through the process of free-electron lasing. Despite the remarkable advances afforded by these facilities, broadening access to x-ray free electron lasers (FELs) would further accelerate scientific progress. A scientific path to broadening access---as opposed to simply building more large-scale accelerator facilities---is to shrink the undulator period. A shorter undulator period reduces the electron energy needed to generate x-ray wavelengths and the distance required for amplification to high powers. To accomplish this, magnetostatic undulators can be replaced by the electromagnetic fields of a laser pulse, where the undulator period is half the laser wavelength and only micrometers in scale compared to centimeters \cite{gallardo1988theory, bonifacio2005quantum, bacci2006transverseEffects,bacciGenesisVsHomemadeCode,SprangleFEL, steiniger2014optical, bonifacio2017design, graves2023cxfel, xu&Mori2024attosecond}. Coupled with the ability to self-seed, these ``laser-driven'' free-electron lasers (LDFELs) have the potential to bring coherent x-ray sources to numerous laser facilities without the need for a coherent seed, a long accelerator, or a large undulator. 

\begin{figure}[ht]
\centering
\includegraphics[scale = 1]{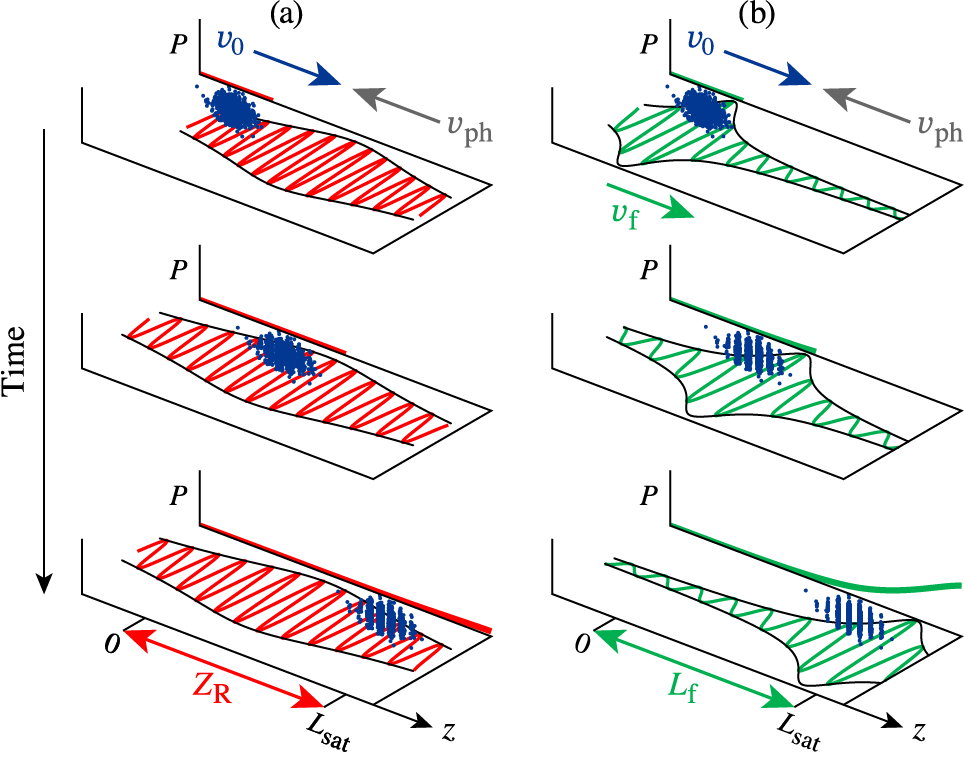}
\caption{Microbunching and x-ray power evolution in a laser-driven free-electron laser (LDFEL). A relativistic electron beam (blue dots) traveling at a velocity $\varv_0$ collides head-on with the phase fronts of a laser pulse traveling at  $\varv_\mathrm{ph} = -c$. (a) A conventional laser pulse with a stationary focus, Rayleigh range $Z_\mathrm{R}$ equal to the saturation length $L_\mathrm{sat}$, and an energy $U = 88 \;\mathrm{J}$  (red). (b) A flying-focus pulse with a moving focus traveling at $\varv_\mathrm{f} = c$, a focal range $L_\mathrm{f} =L_\mathrm{sat}$, and $U = 8 \;\mathrm{J}$ (green). {Both pulses have the same maximum amplitude $a_0$ and wavelength $\lambda_\mathrm{L}$. With the conventional pulse, the longitudinal uniformity of the undulator can only be improved by increasing the spot size and Rayleigh range. With the flying-focus pulse, the peak amplitude travels with the electron beam, allowing the pulse to have a much smaller spot size while still ensuring a longitudinally uniform undulator.} Despite having $11\times$ more energy, the conventional pulse results in a ${\sim}10\times$ lower x-ray power $P$ than the flying focus due to the spatial variation in $a(\bvec{x})$ experienced by the electron beam (see Fig. \ref{fig:2}). Note that the length of the electron beam and period of the microbunches have been elongated for illustrative purposes. 
} \label{fig:1}
\end{figure}

Figure \ref{fig:1}a illustrates a typical LDFEL configuration. A relativistic electron beam collides head-on with the phase fronts of a laser pulse. As the electrons oscillate in the fields of the pulse, they initially undergo inverse Compton scattering and emit incoherent radiation near the wavelength $\lambda_\mathrm{X} = [1+\tfrac{1}{2}a^2(\bvec{x})]\lambda_\mathrm{L}/4\gamma_0^2$, where $\lambda_\mathrm{L}$ is the wavelength of the laser pulse, $a(\bvec{x})$ is the amplitude of its vector potential normalized to $mc^2/e$, $\gamma_0 = (1-\varv_{0}^2/c^2)^{-1/2}$ is the initial electron energy normalized to $mc^2$, and $\varv_0$ is the initial electron velocity. The noise from the incoherent emission seeds a positive feedback loop where the radiation facilitates densification or ``microbunching'' of the electron beam at the length scale $\lambda_\mathrm{X}$. The microbunching in turn enhances the emission, leading to exponential growth of coherent radiation near $\lambda_\mathrm{X}$. The electron trajectories and radiation properties are similar to those in a conventional magnetostatic FEL with an undulator period $\lambda_\mathrm{u} = \lambda_\mathrm{L}/2$ and strength $K = a(\bvec{x})$\cite{SprangleFEL}. The length of the radiation source, however, is highly compressed. For a fixed radiation wavelength $\lambda_\mathrm{X}$, the distance over which the power increases by a factor of $\mathrm{e}$, or gain length, is $L_{\mathrm{g}0} \propto 
\lambda_\mathrm{u}^{5/6}$. Thus, the shortened undulator period---now on the order of microns instead of centimeters---allows for amplification over dramatically shorter distances and the use of significantly lower electron energies.

Despite these advantages, LDFELs face challenges that have, to date, impeded their experimental realization. Foremost among these is that the FEL parameter $\rho \propto \lambda_\mathrm{u}^{2/3}$ of an LDFEL is much smaller than that of a typical magnetostatic FEL. The FEL parameter quantifies the power efficiency, required electron beam quality, and gain bandwidth at saturation. Specifically, $\rho \equiv \gamma_0^{-1}\nu^{1/3}(\lambda_\mathrm{L}a_0/16\pi\sigma_{0})^{2/3}$, where $a_0$ is the maximum normalized vector potential of the laser pulse, $\nu = I_\mathrm{b}\mathrm{\,[A]}/17000$, $I_\mathrm{b}$ is the electron beam current, and $\sigma_{0}$ is the minimum RMS electron beam radius\cite{SprangleFEL}. For typical LDFEL parameters, $\rho = \mathcal{O}(10^{-4})$, which places stringent conditions on the normalized emittance $\epsilon_\mathrm{N}<\sigma_{0}\sqrt{2 \rho}$, energy spread $\Delta \gamma/\gamma_0 < \rho$, and detuning $\Delta \lambda_\mathrm{X}/\lambda_{\mathrm{X}0} \lesssim 2 \rho$ needed for high gain at a target wavelength $\lambda_{\mathrm{X}0}\equiv(1+\tfrac{1}{2}a^2_0)\lambda_\mathrm{L}/4\gamma_0^2$. Satisfying the detuning condition is particularly difficult in an LDFEL because of the spatially varying vector potential: $\Delta \lambda_\mathrm{X} {\sim} \int d\bvec{s} \cdot \nabla a^2(\bvec{x})$, where $\bvec{s}$ is the path of an electron through the undulator. For a conventional laser pulse, this spatial variation is unavoidable. The pulse must be focused to achieve the undulator strengths necessary for high-power x-ray radiation, which introduces both transverse and longitudinal variation (Fig. \ref{fig:1}a). While the spatial uniformity can be improved by increasing the focused spot size (Rayleigh range) with a concomittant increase in the duration, this approach quickly becomes impractical, leading to infeasibly large laser pulse energies for the distances needed to reach peak power, i.e., the saturation length $L_{\mathrm{sat}}$. 

Here we demonstrate that ``flying-focus'' pulses can provide a highly uniform undulator for significantly less laser energy than a conventional laser pulse, enabling the generation of high-power, narrow-bandwidth, coherent x-ray radiation. The flying focus refers to a variety of optical techniques for creating a laser pulse with a time-dependent focal point \cite{sainte2017controlling,froula2018spatiotemporal,palastro2018ionization, palastro2020dephasingless,jolly2020controlling, simpson2022spatiotemporal,ramsey2023exact,ASTRL,Pigeon2024}. The intensity peak formed by the moving focus travels a distance ($L_\mathrm{f}$) far greater than a Rayleigh range while maintaining a near-constant profile. {While previous studies explored the utility of these pulses for generating incoherent x rays from inverse Compton scattering \cite{ramsey2022nonlinear,formanek2022radiation,ye2023enhanced,rinderknecht2024electron,formanek2025enhanced}, this work presents the first application of the flying focus to coherent x-ray generation.} The LDFEL design introduced here employs the ideal flying focus, which creates a moving focal point by focusing a laser pulse through a lens with a time-dependent focal length \cite{simpson2022spatiotemporal,ramsey2023exact}. Within the focal range $L_\mathrm{f}$, the velocity of the resulting intensity peak can be made to travel at $\varv_\mathrm{f} = c$ in the opposite direction of the phase fronts $\varv_\mathrm{ph} = -c$ and in the same direction as the electron beam $\varv_0 \lesssim c$ (Fig. \ref{fig:1}b). {The electrons, both colocated and cotraveling with the intensity peak, experience a nearly uniform undulator strength across the entire focal range.} {Motivated by the wavelength scaling of the FEL parameter, long-wave infrared undulators ($\lambda_\mathrm{L} = 10$ $\mu$m CO$_2$ pulses) are considered}. To compare the x-ray radiation driven by {long-wave} flying focus and conventional pulses, the 3D FEL code \Genesis\cite{reiche1999genesis} was modified to model LDFELs. Simulations employing this code show that with a $\gamma_0 = 35$ electron beam, a flying-focus undulator can produce ${\sim}$1 MW of $\lambda_\mathrm{X} = 2.2$ nm x-ray radiation in only a 1 cm interaction length  using $20\times$ less energy than a conventional laser pulse. Such a source would allow for interrogation of warm dense matter \cite{Hau-Riege2007,Mahieu2018,Falk2018} and falls within the ``water-window'', making it an effective probe for biological matter \cite{spielmann1997generation, berglund2000compact, pertot2017time}.

\section*{Results}
For a conventional laser pulse focused by an ideal lens, the transverse and longitudinal uniformity of the amplitude $a(\bvec{x})$ are characterized by the focused spot size $w_\mathrm{C}$ and Rayleigh range $Z_\mathrm{R} = \pi w_\mathrm{C}^2/\lambda_\mathrm{L}$ (Fig. \ref{fig:1}a). {To avoid spatial detuning and ensure amplification to high powers, the Rayleigh range $Z_\mathrm{R}$ must be longer than the interaction length $L_{\mathrm{int}}$. To sustain the undulator strength over the time it takes the counter-traveling electron beam to traverse the interaction length, the pulse duration must be $T=2L_{\mathrm{int}}/c$.} As a result, the energy of a conventional laser undulator $U_\mathrm{C} \propto a_0^2 w_\mathrm{C}^2 T /\lambda_\mathrm{L}^2 \propto 2a_0^2L_{\mathrm{int}}^2 /\lambda_\mathrm{L}$ scales \textit{quadratically} with the interaction length, where $Z_\mathrm{R} = 2L_{\mathrm{int}}$ has been used. For a flying-focus pulse with an intensity peak that cotravels with the electron beam ($\varv_\mathrm{f} = c$), the transverse and longitudinal uniformity are characterized by the focused spot size $w_\mathrm{FF}$ and focal range $L_\mathrm{f}$ (Fig. \ref{fig:1}b). In this case, the longitudinal uniformity is decoupled from the focused spot size, i.e., $L_\mathrm{f}$ does not depend on $w_\mathrm{FF}$. {To avoid spatial detuning and sustain the undulator strength, the focal range and duration of the flying focus pulse must be $L_\mathrm{f} = L_{\mathrm{int}}$ and $T=2L_{\mathrm{int}}/c$.} Thus, the energy of a flying-focus undulator $U_\mathrm{FF} \propto a_0^2 w_\mathrm{FF}^2 T /\lambda_\mathrm{L}^2 \propto a_0^2 w_\mathrm{FF}^2 L_{\mathrm{int}} /\lambda_\mathrm{L}^2$ scales \textit{linearly} with the interaction length. {The ratio of energies needed to drive an LDFEL with fixed FEL parameter $\rho$, laser wavelength $\lambda_\mathrm{L}$, and radiation wavelength $\lambda_{\mathrm{X}0}$ is then given by
\begin{equation}\label{eq:energyRelation}
   \frac{U_\mathrm{FF}}{U_\mathrm{C}} =  
   \frac{\alpha \pi w_\mathrm{FF}^2}{2\lambda_\mathrm{L} L_\mathrm{int}}.
\end{equation}
}The factor $\alpha {\sim} \mathcal{O}(1)$ is determined by the power ratio of a flying-focus pulse with an arbitrary transverse profile to one with a Gaussian profile of spot size $w_\mathrm{FF}$, both with the same maximum vector potential $a_0$. 

Equation \eqref{eq:energyRelation} elucidates the advantage of using a flying focus to decouple the interaction length from the Rayleigh range. {At the same laser wavelength, the ratio shows that a flying-focus pulse requires less energy than a conventional pulse in LDFEL configurations for which $L_\mathrm{int} > \alpha\pi w_\mathrm{FF}^2/2\lambda_\mathrm{L}$}. When amplifying to saturation, the left-hand side of this condition is determined by the saturation length $L_\mathrm{int} = L_\mathrm{sat}$. Typical saturation lengths tend to be $\mathcal{O}(10)$ times larger than the gain length, i.e.,  $L_\mathrm{sat} = \chi L_\mathrm{g}$, where $\chi {\sim} \mathcal{O}(10)$. The right-hand side of the condition is determined by the electron beam radius $\sigma_0$. To ensure $a(\bvec{x})$ has sufficient transverse uniformity, the spot size of the flying focus should be larger than $\sigma_0$, i.e, $w_\mathrm{FF} = \varpi \sigma_0$, where $\varpi \gtrsim 1$. By combining these scalings and using the conservative approximation that $\alpha \varpi^2 / \chi \approx 1$, the condition can be reexpressed in terms of LDFEL parameters: $(\lambda_\mathrm{L}/\sigma_0)^2>4\pi^2\sqrt{3}\rho$, where the cold beam gain length $L_{\mathrm{g}} = L_{\mathrm{g}0} \equiv \lambda_\mathrm{L}/8\pi\sqrt{3}\rho$ has been used. This condition indicates that the advantage of flying-focus pulses is greater at longer laser wavelengths. {Independent of this advantage, the use of longer laser wavelengths also allows for higher radiated powers and relaxes the requirements on the electron beam quality (see Discussion).} For the parameters considered here $\lambda_\mathrm{L} = 10$ $\mu$m, $L_\mathrm{sat} = 1.03$ cm, and $\alpha \pi w_\mathrm{FF}^2/2\lambda_\mathrm{L} = 0.55$ mm, resulting in  $U_\mathrm{FF}/U_\mathrm{C} = 0.05$---a significant reduction in the required energy when using a flying focus pulse (Table \ref{tab:SimParams}). 

To demonstrate the advantages of an LDFEL with a flying-focus undulator, 3D time-dependent simulations were conducted using the conventional FEL code \Genesis \cite{reiche1999genesis}. The time-dependent model provided by \Genesis allows for seeding from amplified spontaneous emission (SASE) and captures the evolution of the x-ray spectrum. This latter feature is critical for modeling an LDFEL because amplitude variations due to focusing and diffraction can shift the resonant frequency along the interaction length. Despite these features, \Genesis was modified to better model an LDFEL. The equations of motion were updated to include the effects of the Gouy phase, phase-front curvature, plasma dispersion, the spatial profile of $a(\bvec{x})$, and the transverse space-charge repulsion of the electron beam (see Methods).

\begin{table}[htbp]
\centering
\caption{\label{tab:SimParams} Parameters of laser-driven FEL simulations comparing flying-focus and conventional laser undulators. The parameters of the flying-focus pulse were motivated by planned upgrades to the CO$_{2}$ laser at the Brookhaven National Laboratory Accelerator Test Facility\cite{polyanskiy20229, BNL_LaserTable}. In each simulation, the electron beam enters the undulator at $z=0$ with its smallest RMS radius $\sigma_{0}$ and then expands due to the nonzero emittance and space-charge repulsion.}
\begin{tabular}{l l}
\hline
\textbf{LDFEL parameters} \\
\hline
    Target resonant wavelength (nm)             & $\lambda_{\mathrm{X}0} = 2.16$  \\
    Laser wavelength ($\mu$m)                   & $\lambda_\mathrm{L} = 10$\\
    Field amplitude/undulator strength          & $a_0 = 0.35$ \\
    FEL parameter                               & $\rho = 3.09\times10^{-4}$ \\
    1D cold beam gain length (mm)               & $L_{\mathrm{g}0}  = 0.74$ \\ 
    Ideal saturation length (mm)                & $L_{\mathrm{sat}}  = 10.3$ \\ 
\hline
\textbf{Electron beam parameters}  \\ 
\hline                             
Energy  ($mc^2$)                    & $\gamma_0 = 35$ \\
Current (kA)                        & $I_\mathrm{b} = 1$\\   
Minimum RMS radius ($\mu$m)         & $\sigma_{0}=15$ \\
Normalized emittance ($\mu$m-rad)   & $\epsilon_\mathrm{N}\ll  0.37$\\
Energy spread                       & $\Delta\gamma/\gamma_0\ll 3.09\times10^{-4}$\\
\hline
\textbf{Conventional Focus}  \\ 
\hline  
Transverse profile                             & Gaussian \\
Pulse duration (ps)                            & $T =  83$ \\ 
Rayleigh range (cm)                            & $Z_\mathrm{R} =0.11$, $1.25$, $2.50$ \\
Spot size ($\mu$m)                             & $w_\mathrm{C} = 60$, $200$, $283$ \\
Pulse energy (J)                               & $U_{\mathrm{C}} = 8$, $88$, $176$ \\  
\hline
\textbf{Flying Focus}  \\ 
\hline  
 Transverse profile                              & Flattened Gaussian Beam\cite{gori1994flattened}\\%$N=1$ FGB\cite{gori1994flattened} \\
Focal range (cm)                                 & $L_\mathrm{f} = 1.25$ \\  
{Pulse duration (ps)}                            & {$T =  83$} \\ 
%Transform limited pulse duration (ps)            & $\tau = 8.5$ \\ 
%Chirp parameter\cite{palastro2018ionization}     & $\eta = 4.9 $\\  
%Effective focal spot size $(a_0/e)$ ($\mu$m)     & $w_\mathrm{0} = 60$ \\  
Gaussian spot size at focus ($\mu$m)              & $w_\mathrm{FF} = 37.5$ \\  
Gaussian spot at lens (cm)                        & $w_\mathrm{l} = 8.5$ \\  
%F-number                                         & $f/\# = 5.9$\\
Pulse energy (J)                                 & $U_{\mathrm{FF}} = 8$ \\  
\hline 
\end{tabular}
\end{table}

\begin{figure}[htbp]
\centering
\includegraphics[scale = 1]{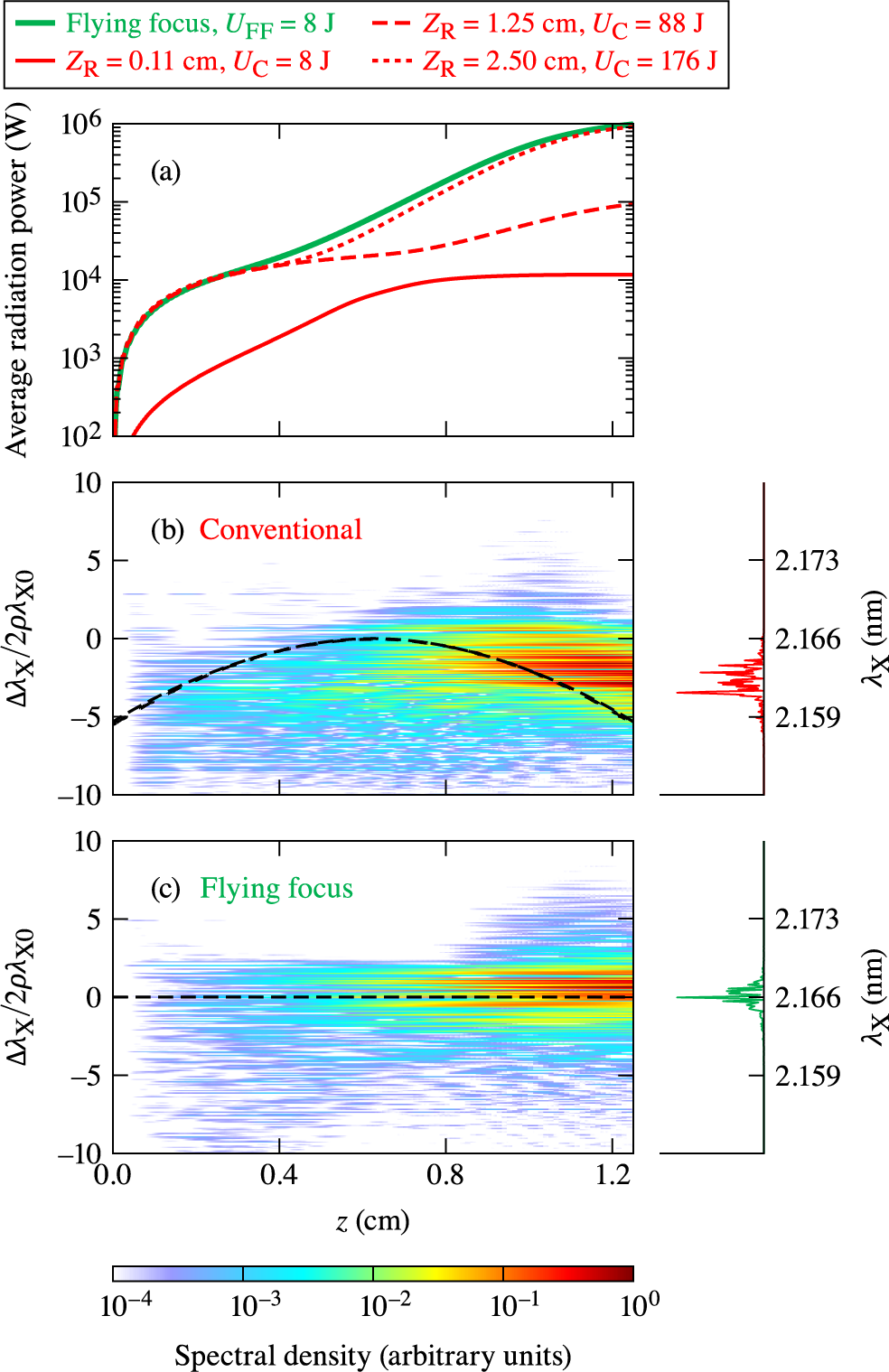}
\caption{Amplification of $\lambda_\mathrm{X} \approx 2.2 \; \mathrm{nm}$ x rays with flying-focus and conventional laser undulators. (a) A flying-focus pulse with $U_\mathrm{FF} = 8$ J  and a conventional pulse with $U_\mathrm{C} = 176$ J drive exponential growth of the x-ray power to saturation, reaching $P = 1$ MW. For the conventional pulse with $U_\mathrm{C} = 8$ J, the power grows rapidly as the electron beam approaches the focal point ($z = 0.625$ cm) and moves into a progressively larger undulator strength, but no microbunching is observed. (b,c) The power spectral density of the x rays produced by the conventional laser undulator with $U_\mathrm{C} = 176$ J and flying-focus undulator with $U_\mathrm{FF} = 8$ J, respectively. The left axes are normalized to the ideal gain bandwidth at saturation, $2\rho$. {The lineouts to the right show the spectra at saturation}. Near saturation, the standard deviation of both spectral densities is ${\sim} 2 \rho$. The dashed line in (b) traces the shift in the resonant wavelength due to amplitude variation in the conventional laser undulator [Eq. \eqref{eq:Con_Res}]. The dashed line in (c) is constant: the resonant wavelength does not shift in the flying-focus undulator. The simulated parameters are provided in Table \ref{tab:SimParams}. }\label{fig:2}
\end{figure}

Figure \ref{fig:2} compares the evolution of the power and spectrum of $\lambda_\mathrm{X} \approx 2.2 \; \mathrm{nm}$ (0.56 keV) x rays amplified in an LDFEL with either a flying-focus or conventional laser undulators. The parameters are displayed in Table \ref{tab:SimParams}. Consistent with the estimate above that $U_\mathrm{FF}/U_\mathrm{C} = 0.05$, the flying-focus pulse required $21\times$ less energy than the conventional pulse to amplify the x rays to the saturated power $P_\mathrm{sat} \approx 1 \; \mathrm{MW}$. In this example, the nonideal effects of a spatially varying undulator strength were isolated from those of electron beam quality by initializing the beam with a negligible normalized emittance ($\epsilon_\mathrm{N}\ll\sigma_{0}\sqrt{2 \rho}$) and energy spread ($\Delta \gamma/\gamma_0 \ll \rho$). 

The conventional laser undulator was able to mitigate spatial detuning and amplify the x rays to saturation with $U_\mathrm{C} = 176\;\mathrm{J}$ of energy, a Rayleigh range $Z_\mathrm{R} = 2 L_\mathrm{sat}$, and a corresponding spot size $w_\mathrm{C} = 283 \; \mu m$.  At a more modest $U_\mathrm{C} = 88\;\mathrm{J}$ and $Z_\mathrm{R} = L_\mathrm{sat}$, the saturated power was $10\times$ lower ($P_\mathrm{sat} \approx 100$ kW). With the same energy as the flying-focus undulator, i.e., $U_\mathrm{C} = 8\;\mathrm{J}$ and $Z_\mathrm{R} = 0.09 L_\mathrm{sat}$, the x-ray power increased rapidly as the electron beam approached the focal point ($z = 0.625$ cm) and encountered larger values of $a(\bvec{x})$, but the radiation was mostly incoherent and microbunching was not observed. In each of these cases, the electron beam was initialized a distance $L_\mathrm{int}/2$ before the focus to optimize the uniformity.

The flying-focus undulator mitigated spatial detuning and amplified the x rays to saturation with only $U_\mathrm{FF} = 8\;\mathrm{J}$ of energy, a focal range $L_\mathrm{f} =  L_\mathrm{sat}$, and a focused spot size $w_\mathrm{FF} = 37.5 \; \mu m$. As opposed to the stationary focus of a conventional laser pulse, the intensity peak of the flying focus moves with the electron beam, ensuring that the electrons experience a nearly uniform and maximum undulator strength $a_0$ over the interaction length. The pulse had a flattened Gaussian transverse profile with $\alpha = 5/2$ (see Methods). {This profile reduces transverse ponderomotive expulsion of the electron beam and provides transverse amplitude uniformity \cite{SprangleFEL,schiavi&BonifacioInvstigates3Deffects}.} The ability to use such a profile is a distinct advantage of the flying focus. With a conventional pulse, focusing and diffraction of a flattened intensity profile would exacerbate the spatial variations of $a(\bvec{x})$. The near-propagation invariance of the flying focus guarantees that the flattened profile persists throughout the focal range. {Simulations (not shown) were also conducted to compare the performance of flying-focus undulators with standard Gaussian and flattened Gaussian profiles. For the same laser-pulse energy and parameters in Table I, the flattened Gaussian profile resulted in an order of magnitude higher x-ray power than the standard Gaussian profile.}

Figure \ref{fig:2}b illustrates how spatial inhomogeneities in the conventional laser undulator due to focusing and diffraction shift the resonant wavelength and modify the x-ray spectrum. Along the propagation axis (i.e., at $r=0$), the amplitude of the conventional pulse is given by $a(z) = a_0/[1+(z-L_\mathrm{int}/2)^2/Z_\mathrm{R}^2]^{1/2}$, where the focal plane is located at $z = L_\mathrm{int}/2$. The resonant x-ray wavelength is then
\begin{equation}\label{eq:Con_Res}
    \lambda_{\mathrm{XC}}(z) = \left[1 + \frac{\tfrac{1}{2}a_0^2}{1+(z-\tfrac{1}{2}L_\mathrm{int})^2/Z_\mathrm{R}^2}\right] \frac{\lambda_\mathrm{L}}{4\gamma_0^2}, 
\end{equation}
where $ \lambda_{\mathrm{XC}}(z=\tfrac{1}{2}L_\mathrm{int}) = \lambda_{\mathrm{X}0}$. Equation \eqref{eq:Con_Res} shows that the resonant wavelength redshifts as the electron beam approaches the focus of the conventional pulse and then blueshifts as the beam moves away from the focus (dashed line in Fig. \ref{fig:2}b). As affirmed by the dashed line in Figs. \ref{fig:2}b, the wavelength shift predicted by Eq. \eqref{eq:Con_Res} is in agreement with the \Genesis simulations. The net effect of the wavelength shifting is that conventional laser undulator does not achieve the maximum radiated power or minimum saturation length of an ideal, monochromatic plane-wave undulator ($P_\mathrm{sat} = 2$ MW and $L_\mathrm{sat} = 1$ cm).

Figure \ref{fig:2}c demonstrates that the longitudinal uniformity of the flying-focus undulator keeps the resonant wavelength tuned to the target wavelength along the entire interaction length (dashed line in Fig. \ref{fig:2}c). However, because the flying-focus pulse has a smaller spot size than the conventional pulse, its ponderomotive force causes a greater transverse expansion of the electron beam. This expansion lowers the beam density. As a result, the flying-focus undulator also does not achieve the maximum radiated power or minimum saturation length of an ideal, monochromatic plane-wave undulator. Nevertheless, both the $U_\mathrm{FF} = 8$ J flying-focus and $U_\mathrm{C} = 176$ J conventional laser undulators produce high-power, narrowband x-ray radiation: at saturation, the standard deviation of both spectral energy densities is $\Delta\lambda_\mathrm{X}/\lambda_{\mathrm{X}0} \approx 2 \rho$.

\begin{figure}[ht]
\centering
\includegraphics[scale = 1]{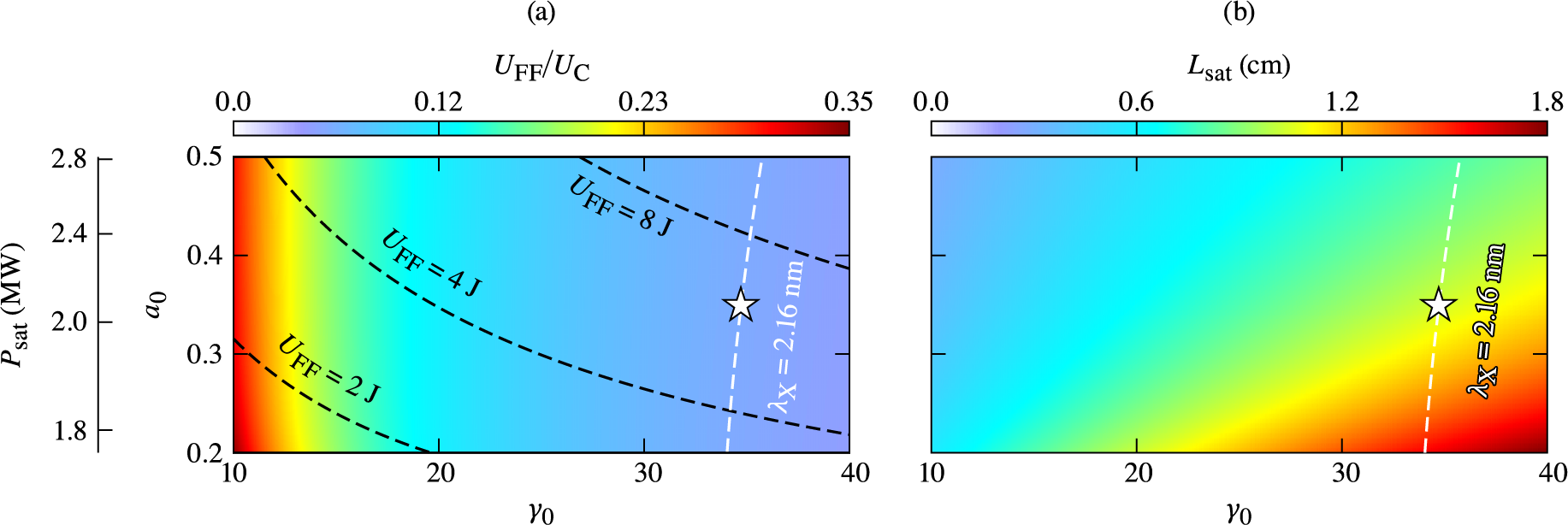}
\caption{The energy-cost benefit of a flying-focus undulator within an LDFEL design space. (a) The ratio of flying focus to conventional pulse energy needed to reach saturation [Eq. \eqref{eq:energyRelation}]. (b) An estimate of the minimum saturation length for each case in (a) with the electron beam current $I$ and width $\sigma_0$ listed in Table \ref{tab:SimParams}. The flying focus provides a larger energy advantage for longer saturation lengths, which coincides with shorter x-ray wavelengths ($\lambda_\mathrm{X} \propto 1/\gamma_0^2$). The second vertical axis (left) shows the maximum achievable power at saturation. The star marks the working point for the simulated examples, and the dashed white line is a curve of constant wavelength $\lambda_\mathrm{X} = 2.16$ nm.}\label{fig:3}
\end{figure}

Figure \ref{fig:3} presents the energy-cost benefit of a flying-focus undulator within a broader LDFEL design space. The energy advantage of the flying focus increases (Fig. \ref{fig:3}a) as the saturation length gets longer (Fig. \ref{fig:3}b), or equivalently, as the target x-ray wavelength gets shorter. The energy ratio is calculated using Eq. \eqref{eq:energyRelation} with $\lambda_\mathrm{L} = 10 \;\mu$m, the focal range of the flying focus set to $L_\mathrm{f} = L_\mathrm{sat}$, the Rayleigh range of the conventional pulse to $Z_\mathrm{R} = 2L_\mathrm{sat}$, and the duration of both pulses to $2L_\mathrm{sat}/c$. Contours of constant wavelength are nearly vertical (e.g., the white dashed line), and the wavelength progressively gets shorter from left to right in each plot ($\lambda_\mathrm{X} \propto 1/\gamma_0^2$). The minimum saturation length was calculated using Eq. (33) of Sprangle \emph{et al.}\cite{SprangleFEL} with the gain length corrected by an empirical factor determined by 3D \Genesis simulations of an ideal plane-wave undulator, i.e, $L_{\mathrm{g}0} \rightarrow 1.3 L_{\mathrm{g}0}$. The use of a plane wave results in a slightly lower energy needed to reach saturation when compared to pulses with transverse structure (cf. Fig. \ref{fig:2}).

The range of $a_0$ values in Fig. \ref{fig:3} was selected to produce a high saturated power (left scale) while ensuring a linear interaction so that the radiation is predominately composed of a single harmonic at $\lambda_\mathrm{X} \approx \lambda_{\mathrm{X}0}$. The maximum achievable saturated power grows with the undulator amplitude $a_0$ but is independent of $\gamma_0$: $P_\mathrm{sat} \approx 0.4\rho \nu \gamma_0 mc^3/r_e \propto a_0^{2/3}$, where $r_e$ is the classical electron radius and the empirical factor of 0.4 is determined by 3D \Genesis simulations of a plane-wave undulator. Note that lower values of $a_0$ reduce the deleterious impact of spatial inhomogeneity on the conventional laser undulator [Eq. \eqref{eq:Con_Res}]. The highest value of $\gamma_0$ was chosen to avoid quantum effects, which increase the classical gain length by a factor $(1+1/\bar{\rho})^{1/2}$, where $\bar{\rho}=  \rho \gamma_0(\lambda_\mathrm{X}/\lambda_\mathrm{AC})$ is the quantum FEL parameter and $\lambda_\mathrm{AC}$ is the Compton wavelength \cite{bonifacio2005quantum,bonifacio2017design}. For all interactions displayed in Fig. \ref{fig:3}, $\bar{\rho} \geq 5$.

\begin{figure}[ht]
\centering
\includegraphics[scale = 1]{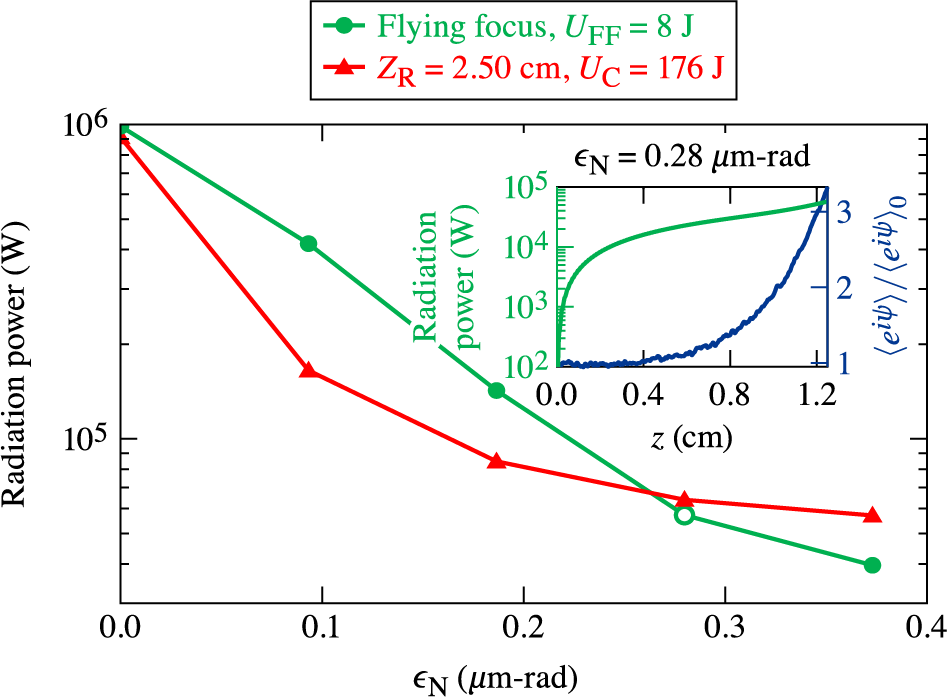}
\caption{The effect of normalized emittance $\epsilon_\mathrm{N}$ on the saturated x-ray power of an LDFEL driven by a flying-focus or conventional laser pulse. The x-ray power drops from $\approx 1$ MW  to a plateau at $\approx 40$ kW as the normalized emittance is increased from $\epsilon_\mathrm{N} = 0$ to $\epsilon_\mathrm{N} = 0.37$ $\mu$m-rad. The initial emittance does not change the fact that a flying-focus pulse (green circles) requires less energy than a conventional pulse (red triangles). The inset displays the evolution of the x-ray power (green) and bunching factor (blue) for a flying-focus undulator with $\epsilon_\mathrm{N} = 0.28$ $\mu$m-rad. All results were obtained from 3D, time-dependent simulations with the LDFEL-modified version of \GenesisNoSpace.} \label{fig:4}
\end{figure}

While the flying focus decreases the laser energy required for a high-gain LDFEL, achieving the necessary electron beam quality remains a formidable challenge. {For the parameters considered in Table \ref{tab:SimParams}, amplification to the maximum saturated power requires an energy spread $\Delta \gamma/\gamma_0 \ll \rho = 3.09\times10^{-4}$ and a normalized emittance $\epsilon_\mathrm{N}\ll \sigma_{0}\sqrt{2 \rho} =  0.37$ $\mu$m-rad---requirements that only become more demanding when attempting to lase at shorter x-ray wavelengths: $\rho \propto \lambda_{\mathrm{X}0}^{1/2}$.} When the electron beam does not satisfy these requirements, the number of electrons that contribute to the instability drops, which reduces the saturated power and elongates the gain length and saturation length \cite{freund1992principles,SprangleFEL}.  Thus, the combination of an imperfect beam and a sufficiently uniform undulator can greatly increase the laser-pulse energy needed for amplification to high powers. 

To assess the impact of imperfect electron beams on the x-ray power generated in the example design (Table \ref{tab:SimParams} and star in Fig. \ref{fig:3}), simulations were run with initial emittances ranging from $\epsilon_\mathrm{N} = 0$ to $\epsilon_\mathrm{N} = 0.37$ $\mu$m-rad. Figure \ref{fig:4} displays the resulting x-ray powers at $z = 1.25$ cm. For both the flying focus and conventional laser undulator, the x-ray power first drops as the initial emittance is increased from $\epsilon_\mathrm{N} = 0$ to $\epsilon_\mathrm{N} = 0.28$ $\mu$m-rad and then plateaus to the incoherent power at emittances $\epsilon_\mathrm{N} > 0.37$ $\mu$m-rad. At lower emittances, better amplitude uniformity makes the flying-focus undulator more resilient to the adverse effects of beam quality. At larger emittances, the x-ray power produced by the $U_\mathrm{C} =176$ J conventional laser undulator is slightly greater than that of the $U_\mathrm{FF} =8$ J flying-focus undulator. This is because the flying-focus pulse has a smaller spot size: when $\epsilon_\mathrm{N} > 0$, a larger portion of electrons have sufficient transverse momentum to laterally escape the fields of the flying-focus pulse. {Simulations (not shown) were also run to confirm that an energy spread $\Delta \gamma/\gamma_0 = \rho = 3.09\times10^{-4}$ degrades the saturated power by nearly the same amount as an emittance $\epsilon_\mathrm{N} =\sigma_{0}\sqrt{2 \rho} = 0.37$ $\mu$m-rad}. In either case, imperfect electron beams do not change the fact that the flying-focus pulse requires much less energy than the conventional pulse to obtain a comparable radiation power.

The inset in Fig. \ref{fig:4} shows the evolution of the radiated power and bunching factor through the flying-focus undulator for the electron beam with $\epsilon_\mathrm{N} =  0.28$ $\mu$m-rad = $\tfrac{3}{4}\sigma_{0}\sqrt{2 \rho}$. The bunching factor $\langle e^{i\psi}\rangle$, where $\psi$ is the ponderomotive phase and $\langle \rangle$ denotes an average over all electrons, is normalized to its initial value at $z=0$. While amplification of the x-ray power and exponential growth of the bunching factor are observed at this emittance, the maximum x-ray power is reduced by an order of magnitude compared to the ideal electron beam ($\epsilon_\mathrm{N} = 0$). Note that amplification in the range $0.09$ $\mu$m-rad $\leq\epsilon_\mathrm{N} \leq 0.28$ $\mu$m-rad violates the Pellegrini criterion for spatial overlap between the electron beam and x-ray pulse \cite{pellegrini1988progress}. The criterion states that the distance over which the electron beam spreads transversely, i.e., $\beta^* = \gamma_0 \sigma_0^2/\epsilon_\mathrm{N}$, should be greater than the Rayleigh range of the x-ray pulse: $\epsilon_\mathrm{N} < \gamma_0 \lambda_\mathrm{X}/4\pi =0.006$ $\mu$m-rad. However, in an LDFEL, both $\beta^*$ and the x-ray Rayleigh range are longer than the saturation length. Thus, the electron beam remains spatially overlapped with the x-ray beam by virtue of the short saturation length. The next most stringent requirement on the emittance limits emittance-induced spectral broadening: $\epsilon_\mathrm{N}<\sigma_{0}\sqrt{2 \rho}$ \cite{bacci2006transverseEffects, steiniger2014optical, bonifacio2017design}. This condition is consistent with the absence of amplification observed at $\epsilon_\mathrm{N} = \sigma_{0}\sqrt{2 \rho} =  0.37$ $\mu$m-rad, where the final bandwidth $\Delta \lambda_\mathrm{X}/\lambda_{\mathrm{X}0} \approx 7 \rho$ was much greater than $2\rho$. 
%checked x2

{In addition to the requirements on the emittance and energy spread, the electron beam must be aligned with the propagation axis of the laser undulator. A lateral displacement $|d|<w$, assuming $\sigma_0 < w$, and angle between the electron beam and optical axis $\Theta\lesssim w /L_\mathrm{sat}$ ensure that the electron beam starts and stays within the spot size of the laser pulse as it traverses a saturation length. For the simulated parameters (Table 1), $|d_\mathrm{FF}|<37.5$ $\mu$m and $\Theta_{\mathrm{FF}}\lesssim  0.003 $ whereas $|d_\mathrm{C}|< 283$ $\mu$m and $\Theta_{\mathrm{C}} \lesssim 0.023$. While the conventional pulse has larger tolerances, they come at the cost of $20\times$ more laser energy. (the spot size of the flying focus could also be increased to improve its tolerances at the cost of more laser energy). Furthermore, the estimated tolerances for the conventional pulse are optimistic: The conventional pulse has a transverse Gaussian profile, which introduces additional amplitude variation and detuning. The flying focus, on the other hand, can have a flattened Gaussian transverse profile with minimal amplitude variation and detuning within the spot size.} {Regardless of whether one opts for a flying-focus or conventional laser undulator, it will be challenging to satisfy the strict requirements on the electron-beam quality and alignment when seeding from noise. As an alternative, the electron beam can be pre-bunched, which can relax or even circumvent these requirements \cite{graves2023cxfel,GravesPreBunch,xu&Mori2024attosecond}.} 

\section*{Discussion}
The ratio of energies expressed in Eq. \eqref{eq:energyRelation} demonstrates that a flying-focus pulse decreases the energy needed for a uniform undulator when the interaction length exceeds the Rayleigh range. The appearance of $\lambda_\mathrm{L}$ in the denominator indicates that this benefit is reduced for shorter laser wavelengths and, by extension, shorter undulator periods. This is because for the same spot size, shorter-wavelength laser pulses have longer Rayleigh ranges, which provide better amplitude uniformity. Moreover, the gain and saturation lengths are smaller at shorter wavelengths, which means the interaction length $L_\mathrm{int}$ is smaller. This suggests an even further reduction in the benefit of the flying focus at shorter wavelengths: $U_\mathrm{FF}/U_\mathrm{C} \propto 1/ \lambda_\mathrm{L} L_\mathrm{int}$. 

An LDFEL designed for longer laser wavelengths and undulator periods has several advantages that incentivize working in a regime where the flying focus provides an energy savings. The important parameters for designing an LDFEL are the FEL parameter $\rho$, gain length $L_{\mathrm{g}0}$, radiation power at saturation $P_\mathrm{sat}$, electron beam energy $\gamma_0$, and quantum FEL parameter $\bar{\rho}$. Consider two undulator periods, $ \lambda_1= \lambda_{L1}/2 $ and $\lambda_2= \lambda_{L2}/2$. For the same target x-ray wavelength, beam current, beam radius, and peak vector potential, the ratios of the design parameters are given by
\begin{equation}\label{eq:comparisions}
        \frac{\rho_1}{\rho_2} = \left(\frac{\lambda_1}{\lambda_2}\right)^{1/6}, \hspace{20pt} 
        \frac{L_{\mathrm{g}0,1}}{L_{\mathrm{g}0,2}} = \left(\frac{\lambda_1}{\lambda_2}\right)^{5/6}, \hspace{20pt}  
        \frac{P_{\mathrm{sat},1}}{P_{\mathrm{sat},2}} = \left(\frac{\lambda_1}{\lambda_2}\right)^{2/3},
        \hspace{20pt} 
        \frac{\gamma_{0,1}}{\gamma_{0,2}} = \left(\frac{\lambda_1}{\lambda_2}\right)^{1/2},
        \hspace{20pt} 
        \frac{\bar{\rho}_1}{\bar{\rho}_2} = \left(\frac{\lambda_1}{\lambda_2}\right)^{2/3}.
\end{equation}
These ratios demonstrate that undulators with longer periods have higher efficiencies, relax the requirements on the beam quality, result in higher saturated powers, and suffer less degradation due to quantum effects. Undulators with shorter periods, on the other hand, allow for smaller interaction lengths and lower electron energies. As an example, comparing a typical glass laser with $\lambda_1 = 0.5$ $\mu$m to a CO$_2$ laser with $\lambda_2 = 5$ $\mu$m yields
$\rho_{1}/\rho_{2}= 0.68$, $L_{\mathrm{g}0,1}/L_{\mathrm{g}0,2}= 0.15$, $P_{\mathrm{sat},1}/P_{\mathrm{sat},2}= 0.21$, $\gamma_{0,1}/\gamma_{0,2} = 0.32 $, and $\bar{\rho}_{1}/\bar{\rho}_{2}= 0.21$. Aside from the advantages of longer undulator periods, imperfect electron beams and quantum effects can substantially increase the gain and interaction lengths, which may make the flying focus energetically favorable even when using shorter-wavelength lasers. This will be a topic of future investigation. {Note that the scalings appearing in Eq. \eqref{eq:comparisions} are independent of undulator type and can be used to compare LDFELs to magnetostatic FELs or other schemes, such as plasma-based  \cite{rykovanov2015plasma} or crystal undulators \cite{korol1998coherent, bellucci2004crystal}.} The main advantages of LDFELs are the smaller distances required to reach saturation and the much lower electron energies needed for the same radiation wavelength. 

In the flying focus configuration implemented here, the electron beam collides head-on with the phase fronts and cotravels with the moving intensity peak [Fig. \ref{fig:1}(b)]. The velocity control and amplitude uniformity afforded by the flying focus allows for other interaction geometries as well. The ideal flying focus uses a lens with a time-dependent focal length to produce a focal point that moves longitudinally, either parallel or antiparallel to the phase fronts \cite{simpson2022spatiotemporal,ramsey2023exact}. With the addition of a time-dependent tilt, the focal point could move in both the transverse and longitudinal directions while the phase fronts move only in the longitudinal direction\cite{gong2024laser}. As before, the velocity of the focal point can be preset to $\varv_\mathrm{f} \lesssim c$, so that the electron beam cotravels with the intensity peak and experiences a uniform amplitude over an extended distance. This configuration complements that proposed in Steiniger \emph{et al.}\cite{steiniger2014optical} and Debus \emph{et al.}\cite{Debus2010} where pulse-front tilt is used to ensure amplitude uniformity throughout the interaction. The difference is that with the flying focus the \textit{focal point} moves with the electron beam. In both cases, the undulator period is increased from $\lambda_\mathrm{L}/2$ to $\lambda_\mathrm{L}/[1-\cos({\vartheta})]$, where $\vartheta$ is the angle between the electron velocity and the phase velocity of the laser pulse. The flexibility to adjust $\vartheta$ allows for optimization of the undulator period. Specifically, one can make the replacement $\lambda_1/\lambda_2 \rightarrow 2/[1-\cos({\vartheta})]$ in Eq. \eqref{eq:comparisions}.

{Alternative methods for improving the longitudinal uniformity of an LDFEL undulator include the use of a plasma channel \cite{Durfee1993, Clark2000,pogorelsky2004experiments} or a Bessel beam. A plasma channel acts as a waveguide that circumvents diffraction by confining a laser pulse transversely \cite{Durfee1993}. In principle, a laser pulse can propagate through the waveguide while maintaining a fixed transverse profile. However, this requires injection of a pulse with flat phase fronts and a profile that is matched to a transverse mode of the waveguide. If these requirements are not met, the pulse will propagate as a superposition of waveguide modes with unique phases that evolve longitudinally\cite{Clark2000}. The resulting interference would spoil the uniformity of the undulator and shift the resonant wavelength along the channel. To ensure a wavelength detuning of no more than several $\rho$ for the parameters considered here, the incident spot size would have to be matched to that of transverse mode with sub-micron precision.} {Bessel beams, i.e., laser beams ``focused'' by axicon lenses, feature an intensity profile that is localized transversely but is nearly uniform longitudinally. Unlike a lens which concentrates rays that start from different radii to a single point, an axicon distributes rays from different radii to different longitudinal locations. For an axicon designed to distribute the rays across the interaction length of an LDFEL $L_\mathrm{int}$, the required energy in the Bessel beam is $U_\mathrm{B}\propto L_\mathrm{int}^3$ and the ratio of energies required in a flying focus and Bessel beam is $U_\mathrm{FF}/U_{\mathrm{B}} = \alpha w_\mathrm{FF}^2/2L_\mathrm{int}^2$. Thus, the flying focus requires less energy than the Bessel beam when $w_{\mathrm{FF}} <\sqrt{2}L_\mathrm{int}/\alpha$, which would always be the case. For the parameters in Table 1, $U_\mathrm{FF}/U_{\mathrm{B}} \approx 10^{-5}$.}

The use of the ideal flying focus was motivated by its conceptual simplicity. Other optical techniques could also be used to achieve the LDFEL configuration depicted in Fig. \ref{fig:1}b. These include “Arbitrarily Structured Laser Pulses” (ASTRL pulses)\cite{ASTRL} and the experimentally demonstrated “space-time wave packets” \cite{kondakci2019optical,yessenov2022space} or ``chromatic flying focus'' \cite{froula2018spatiotemporal,jolly2020controlling}. An LDFEL based on the ``chromatic'' flying focus was also simulated as a part of this study. The chromatic flying focus creates a moving focal point by focusing a chirped laser pulse with a chromatic lens. The chromatic lens focuses each frequency to a different location within an extended focal range, while the chirp determines the arrival time of each frequency at its focus. The simulations (not shown) revealed that the variation in the undulator period caused by the chirp can modify the x-ray spectrum and reduce the saturated power. Laser pulse propagation simulations of the chromatic flying focus, following the method outlined in Palastro \emph{et al.}\cite{palastro2018ionization}, indicate that the focal geometry can be adjusted to reduce to chirp and achieve the same saturated power as the ideal flying focus. These modifications to the focal geometry require 37.5 J of laser energy to amplify the x-ray radiation to 1 MW in 1.25 cm compared to the 176 J needed with a conventional laser pulse.

In conclusion, flying-focus pulses can substantially reduce the energy required to produce coherent, narrowband, high-power x-rays in a laser-driven free-electron laser. In contrast to the static focal point of a conventional laser pulse, the dynamic focal point of a flying-focus pulse travels with the electron beam, ensuring a uniform undulator over the entire interaction length. Simulations of a design based on CO$_2$ laser parameters showed that a flying focus can produce ${\sim}$1 MW of $\lambda_\mathrm{X} = 2.2$ nm x-ray radiation from a 17.5 MeV electron beam in a 1.25 cm interaction length using $20\times$ less energy than a conventional laser pulse. While electron beam quality does affect the final x-ray power, it does not change the fact that the flying focus provides an energy advantage. The velocity control and extended interaction lengths at high intensity enabled by the flying focus provide a path to LDFELs with currently achievable laser energies.

\section*{Methods}
% SEE Equations of Motion document for my og derivation p_rapid is not p_rapid in those notes
\subsection*{Laser-driven free-electron laser model}
In a laser-driven free-electron laser (LDFEL), the electromagnetic field of a laser pulse with a wavelength $\lambda_\mathrm{L} {\sim} \mathcal{O}(1 \,\mathrm{to}\, 10 \; \mu\mathrm{m})$ provides an undulator that allows for the emission and amplification of electromagnetic fields at other wavelengths. The laser pulses considered here propagate in the negative $\unitvec{z}$ direction and are circularly polarized. The electromagnetic fields of the pulses are modeled in terms of the vector potential
\begin{equation}
    \bvec{a}_\mathrm{L} = \frac{a_\mathrm{L}(\bvec{x},t)}{\sqrt{2}}\left[\cos\big(k_\mathrm{L} z+\omega_\mathrm{L}t+ \phi(\bvec{x},t)\big)\unitvec{x}+\sin\big(k_\mathrm{L} z+\omega_\mathrm{L}t+\phi(\bvec{x},t)\big)\unitvec{y}\right],
\end{equation}
where $\omega_{L} = 2 \pi c/\lambda_\mathrm{L}$, $ck_\mathrm{L} \equiv (\omega_\mathrm{L}^2 -\omega_\mathrm{pl}^2/\gamma_0)^{1/2}$, $\omega_\mathrm{pl} =c \sqrt{2\nu}/\sigma_{0}$ is the plasma frequency of the electron beam, and potentials are normalized to $mc^2/e$ throughout. The amplitudes $a_\mathrm{L}$ and phases $\phi$ are real quantities and are defined in the Laser Pulse Model subsection. A head-on collision between a relativistic electron beam and these fields results in the emission and amplification of a circularly polarized x-ray pulse (Fig. \ref{fig:1} and Table \ref{tab:SimParams}) that propagates in the positive $\unitvec{z}$ direction. The electromagnetic fields of the x-ray pulse are modeled with the vector potential 
\begin{equation}
        \bvec{a}_\mathrm{X} = \frac{a_\mathrm{X}(\bvec{x},t)}{2}\exp\left[i(k_\mathrm{X} z-\omega_\mathrm{X}t)\right]\unitvec{e}_\perp +\mathrm{c.c.},\\
\end{equation}
where $\omega_\mathrm{X} = 2\pi c/\lambda_\mathrm{X} = ck_\mathrm{X}$, $\unitvec{e}_\perp = (\unitvec{x}+i\unitvec{y})/\sqrt{2}$, and $a_\mathrm{X}$ is complex. The plasma dispersion contribution to $k_\mathrm{X}$ is neglected because $\omega_\mathrm{X}\gg \omega_\mathrm{pl}$. 

The motion of beam electrons in the laser and x-ray pulses can be separated into rapid and slowly varying components. The rapid motion describes the oscillations in the fields of the laser pulse, while the slow motion describes trajectory modifications due to ponderomotive forces and the space charge fields. The rapid oscillations modulate the slow ``guiding-center'' evolution of the positions and momenta. In the following, a tilde (${\sim}$) is used to distinguish the rapidly varying momenta from the unadorned guiding-center momenta. 

The rapidly varying momenta are equal to the local value of the laser-pulse pulse vector potential: $\tilde{\bvec{p}}= \bvec{a}_\mathrm{L}$, where the momenta have been normalized by $mc$. The guiding-center dynamics are governed by the Hamiltonian
\begin{equation}\label{eq:Ham}
    H = \left[1+\left({P}_z + a_{\mathrm{SC}}({\bvec{{x}}})\right)^2+|{\bvec{p}}_\perp|^2 + \varphi_{\mathrm{P}}({\bvec{x}},t)\right]^{1/2} - \varphi_{\mathrm{SC}}({\bvec{x}}),
\end{equation}
where
\begin{equation}\label{eq:pondpot}
    \varphi_{\mathrm{P}}({\bvec{x}},t) =  \tfrac{1}{2}a_\mathrm{L}^2({\bvec{x}},t) + \tfrac{1}{2}a_\mathrm{L}({\bvec{x}},t)\left[a_\mathrm{X}({\bvec{x}},t)\exp({i\psi})+\mathrm{c.c}\right]
\end{equation}
is the ponderomotive potential,
\begin{equation}\label{eq:pondphase}
    \psi = (k_\mathrm{X}+k_\mathrm{L}){z} - (\omega_\mathrm{X} - \omega_\mathrm{L})t +\phi({\bvec{x}},t), 
\end{equation}
is the phase of the ponderomotive beat, $\varphi_{\mathrm{SC}}$ and $a_{\mathrm{SC}}$ are the scalar and vector potentials describing the space-charge fields, and ${P}_z = {p}_z -a_{\mathrm{SC}}$ is the longitudinal canonical momentum. In general, $|a_\mathrm{L}| \gg |a_\mathrm{X}|$, thus terms $\propto|a_\mathrm{X}|^2$ have been neglected in $H$. The derivation of $H$ involves a cycle average over the period of the laser pulse in a frame moving with the electron beam ($z \approx \varv_0 t$). The ponderomotive phase $\psi$ varies slowly in this frame because $[(k_\mathrm{X}+k_\mathrm{L})\varv_0 - (\omega_\mathrm{X} - \omega_\mathrm{L})] \approx 0$ or, equivalently, $\omega_\mathrm{X}\approx 4\gamma_0^2 \omega_\mathrm{L}$.

The equations of motion for the guiding-center coordinates, momenta, and energy are derived from the Hamiltonian. The coordinates evolve according to
\begin{equation}
\frac{d\bvec{x}}{dt} = c\frac{\partial H}{\partial \bvec{ p}} = \frac{c\bvec{p}}{\gamma},
\end{equation}
where $\gamma = (1+|{\bvec{p}}|^2 + \varphi_{\mathrm{P}})^{1/2}$. 
The transverse momenta evolve in response to the transverse ponderomotive and space-charge forces
\begin{equation}
    \frac{d{\bvec{p}}_\perp}{dt} = -c\nabla_\perp H = -\frac{c}{4\gamma}\nabla_\perp a_\mathrm{L}^2 + c\nabla_\perp \varphi_{\mathrm{SC}} -\frac{c {p}_z}{\gamma}\nabla_\perp a_{\mathrm{SC}}.
\end{equation}
Here, $| a_\mathrm{L}^2| \gg 2|a_\mathrm{L} a_\mathrm{X}|$ has been used to drop terms ${\propto}a_\mathrm{L} a_\mathrm{X}$ in the transverse ponderomotive force. The work done by the ponderomotive and space-charge forces modifies the electron energy
\begin{equation}
    \frac{d{\gamma}}{dt} = \frac{d(H + \varphi_\mathrm{SC})}{dt} = \partial_t H + \left(\partial_t + \frac{{c\bvec{p}}}{{\gamma}}\cdot\nabla \right)\varphi_\mathrm{SC} = 
    -\frac{1}{4{\gamma}}\omega_\mathrm{X} a_\mathrm{L}\left[ia_X \exp{(i\psi)}+\mathrm{c.c.}\right] + \frac{{\bvec{p}}}{{\gamma}}\cdot \nabla \varphi_{\mathrm{SC}},
\end{equation}
where $\omega_\mathrm{X}\gg\omega_{L}$ has been used in the time derivative of $\varphi_{\mathrm{P}}$ and it has been assumed that $|\partial_t a_\mathrm{L}^2| \ll |\omega_\mathrm{X} a_\mathrm{L} a_\mathrm{X}|$. This latter condition is always satisfied for conventional laser pulses with flattop temporal profiles, such as those used in the simulations. For the flying-focus pulses of interest, where the intensity peak is colocated and cotravels with the electron beam, it is convenient to recast the condition in terms of the coherent x-ray power: $P_X \,[\mathrm{W}]\gg 3.5\times10^5 (a_0 \sigma_{0} L_\mathrm{b}\lambda_\mathrm{L}^2/w_\mathrm{FF}^4)^2$. Evaluating the right-hand side with the parameters in Table \ref{tab:SimParams} and $L_\mathrm{b} = 50 \; \mu$m yields $P_X \,[\mathrm{W}]\gg 60$, which is easily satisfied.

In \GenesisNoSpace, the equations of motion are integrated in $z$ instead of $t$. With the substitution $\tfrac{d}{dt} =  \varv_z\tfrac{d}{dz} $, the full set of guiding-center equations of motion becomes
\begin{align}
        &\frac{d\bvec{{p}}_\perp}{d {z}} = -\frac{1}{4{p}_z} \nabla_\perp a_\mathrm{L}^2({\bvec{x}})+ \frac{{\gamma}}{{p}_z}\nabla_\perp \varphi_{\mathrm{SC}}({\bvec{x}}) -\nabla_\perp a_{\mathrm{SC}}({\bvec{x}}),\label{eq:laserPond}\\
        &\frac{d\bvec{{x}}_\perp}{d {z}} = \frac{\bvec{{p}}_\perp}{{p}_z},\\
        &\frac{d{\gamma}}{d {z}} = -\frac{1}{4{p}_z}k_\mathrm{X}a_\mathrm{L}(\bvec{x})\left[ia_X(\bvec{x},t) \exp{(i\psi)}+\mathrm{c.c.}\right]+\frac{{\bvec{p}}}{{p}_z}\cdot \nabla \varphi_{\mathrm{SC}}({\bvec{x}})\label{eq:EOM_start},\\
        &\frac{d\psi}{d {z}} = (k_\mathrm{X}+k_\mathrm{L}) - \frac{1}{{\varv}_z}(\omega_\mathrm{X} - \omega_\mathrm{L})+\frac{d\phi({\bvec{x}})}{d {z}},\label{eq:pondPhaseEvolve}\\
        &{p}_z = \left[\gamma^2 - (1 + |{\bvec{p}}_\perp|^2 + \varphi_\mathrm{P})\right]^{1/2}\label{eq:EOM_end},
\end{align}
where $t \approx {z}/c$ has been used in $\phi$. This set of equations is similar to the set describing electron motion in a conventional magnetostatic FEL with an undulator period $\lambda_\mathrm{u} = \lambda_\mathrm{L}/2$ (cf. Reiche\cite{reiche2000numerical}). There are, however, two important distinctions. First, the focusing geometry used for the laser pulse contributes a phase $\phi(\bvec{x})$ that can spatially detune the FEL instability [Eq. \eqref{eq:pondPhaseEvolve}]. Second, the transverse ponderomotive force [first term in Eq. \eqref{eq:laserPond}] pushes electrons from regions of high to low undulator strength. When the intensity of the laser pulse is peaked on-axis, this has the opposite effect of the natural focusing that occurs in a magnetostatic undulator.

The motion of the electron beam in the laser undulator results in a transverse current that drives the x-ray radiation. The envelope of the x-ray pulse $a_\mathrm{X}$ evolves according to the paraxial wave equation
\begin{equation}\label{eq:parX}
    \left[\nabla_\perp^2  + 2 i k_\mathrm{X} \left(\frac{\partial }{\partial z}+\frac{\partial }{c\partial t}\right)\right]a_\mathrm{X} =  4\pi r_e \sum_j \frac{a_{L}({\bvec{x}}_j)}{{\gamma}_j}\delta(\bvec{x} - {\bvec{x}}_j)e^{-i\psi_j},
\end{equation}
where $r_e $ is the classical electron radius and the summation is over all electrons. In practice, \Genesis uses ``macro''-electrons to avoid simulating all of the electrons present in an actual beam. To solve Eqs. \eqref{eq:laserPond}--\eqref{eq:parX}, the macroelectrons are initialized in $t$ as ``slices'' of duration $\lambda_{\mathrm{X}0}/c$, where $\lambda_{\mathrm{X}0}$ is the target radiation wavelength \cite{reiche1999genesis,reiche2007recent}. The macroelectron motion and x-ray envelope $a_\mathrm{X}$ are advanced in steps of $\Delta z$. After a specified number of longitudinal steps $N_z$, the envelope is advanced in time by $\Delta t = N_z\lambda_{\mathrm{X}0}/c$. This numerical approach is valid when $c\Delta t$ is much shorter than the cooperation length, $L_\mathrm{c} = \lambda_{\mathrm{X}0}/4\pi\rho$, which defines the slippage of the x-ray pulse relative to the electron beam over a gain length. The underlying assumption is that the instability does not grow significantly in the time it takes the x-ray radiation to ``slip'' by one electron slice (i.e., by a length $\lambda_{\mathrm{X}0}$). 

The density and current of the electron beam also produce space-charge fields that feedback onto the motion. The $\varphi_{\mathrm{SC}}$ and $a_\mathrm{SC}$ terms appearing in Eqs. \eqref{eq:laserPond} and \eqref{eq:EOM_start} correspond to the Lorentz forces from the longitudinal and radial electric fields ($\bvec{E} = - \nabla \varphi_\mathrm{SC}$) and azimuthal magnetic field ($B_{\theta} = -\partial_r a_\mathrm{SC}$, where $r = |\bvec{x}_{\perp}|$). The longitudinal electric field is assumed to be periodic with respect to the ponderomotive phase, i.e., $E_z = \tfrac{1}{2}\sum_\ell \hat{E}_{z,\ell}e^{i\ell\psi}+\mathrm{c.c}$, where the amplitudes ($\hat{E}_{z,\ell}$) are real. Substitution of the Fourier series into the inhomogeneous wave equation provides an equation for each amplitude \cite{tran1990review,reiche2000numerical}
\begin{equation}\label{eq:EZ}
    (\nabla_\perp^2 - 4k_\mathrm{L}k_{\mathrm{X}0})\hat{E}_{z,\ell} =  -8r_ek_\mathrm{L}k_{\mathrm{X}0}\ell \sum_j \delta({\bvec{x}}_\perp-{\bvec{x}}_{\perp,j})\sin(\ell \psi_j),
\end{equation}
where the summation is over all macroelectrons in a time slice and $\delta(z-z_j) =  k_{\mathrm{X}0}\delta(\psi - \psi_j)$ has been used. The approximation $z \approx ct$ has also been used to approximate and $\partial_t \rho \approx -\partial_z J_z$, where $\rho$ and $J_z$ are the charge and longitudinal current densities, respectively. The simulations presented above include the $\ell = 1$ mode of the longitudinal electric field. 

A laser-based undulator lowers the electron energy needed to produce x rays, but operating at lower energies exacerbates the effects of transverse space-charge repulsion. As a rough estimate, transverse space-charge forces will cause a significant increase in the electron beam radius over a length $L_\mathrm{SC} \equiv 2 c\gamma_0^{3/2}/\omega_\mathrm{pl}$. For the presented design (Table \ref{tab:SimParams}), this length is comparable to the interaction length: $L_\mathrm{SC} = 1.8$ cm and $L_\mathrm{int} = 1.25$ cm. To capture the effect of these forces, a self-consistent calculation of the transverse space-charge fields ($E_r$ and $B_\theta$) and their feedback onto the electron motion was added to \GenesisNoSpace. As with $E_z$, $E_r$ and $B_\theta$ are assumed to be periodic with respect to the ponderomotive phase. Unlike $E_z$, however, these field components are nonzero for $\ell = 0$. Assuming cylindrical symmetry, the $\ell = 0$ amplitudes of the transverse space-charge fields satisfy
\begin{align}
    \frac{\partial}{\partial r}(r \hat{E}_{r,0}) &= -\frac{r_e k_{\mathrm{X}0}}{\pi}\sum_j \delta(r-r_j),\\
   \frac{\partial}{\partial r}(r \hat{B}_{\theta,0}) &= -\frac{r_e k_{\mathrm{X}0}}{\pi}\frac{\varv_{0}}{c}\sum_j \delta(r-r_j).
\end{align}
The same approximations applied to the delta functions in Eq. \eqref{eq:EZ} are applied here. In addition, the longitudinal velocity in the summation for $\hat{B}_{\theta,0}$ is approximated by the initial velocity ($\varv_{z,j} \approx \varv_0$), such that $\hat{E}_{r,0} = \varv_{0}\hat{B}_{\theta,0}/c$. Note that the $\ell = 0$ term for the space-charge fields is the dominant contribution; the $\ell = 1$ term is $\mathcal{O}(\lambda_{\mathrm{X}0}^2/\sigma_0^2)$ smaller.

One of the most-striking differences between conventional FEL and LDFEL simulations is the requirement on the transverse resolution. A series of simulations using the LDFEL-modified version of \Genesis was conducted to determine the transverse resolution required for convergence of the saturated power and saturation length. For simplicity, an ideal, plane-wave laser undulator was considered. The simulations confirmed the analytic calculations in Sprangle \emph{et al.}\cite{SprangleFEL} that at a minimum it is necessary to resolve transverse wave numbers up to
\begin{equation}
    k_{\perp,\mathrm{min}} \approx 3\sqrt{2}\left(\frac{k_{\mathrm{X}0}}{L_{\mathrm{g}0}}\right)^{1/2}.
\end{equation}
Because the gain length ($L_{\mathrm{g}0}$) is much smaller in an LDFEL, the LDFEL simulations require a much higher transverse resolution to accurately model all of the amplified wavevectors: $\Delta x < \Delta x_\mathrm{max} \approx \pi/k_{\perp,\mathrm{min}} \propto L_{\mathrm{g}0}^{1/2}$.
%checkedx2

{The high transverse resolution required to model LDFEL amplification from noise in \Genesis make a calculation of the x-ray diffraction angle at saturation computationally challenging. The divergence angle at saturation $\theta_\mathrm{D}$ can, however, be bounded: $2\sqrt{2}/k_{\mathrm{X}0}\sigma_0<\theta_\mathrm{D}<\sigma_0/\sqrt{2}L_\mathrm{sat}$. For the simulations presented above, $6.5\times10^{-5}<\theta_\mathrm{D}<8.5\times10^{-4}$. The lower bound assumes the x-ray pulse is diffraction limited with a minimum spot size $\sigma_0/\sqrt{2}$. The upper bound assumes the maximum transverse wavevector exits the electron beam after a saturation length. Future work will pursue modifications to the \Genesis output routines to better manage the highly resolved x-ray field data.}

\subsection*{Laser pulse model}
The amplitudes $a_\mathrm{L}$ and phases $\phi$ of the conventional and flying-focus pulses used in the \Genesis simulations are given by
\begin{align}
        a_\mathrm{C}({\bvec{x}}) &= a_0\frac{w_\mathrm{C}}{w(\zeta)}\mathrm{exp}\left[ -\frac{r^2}{w(\zeta)} \right],\\
        \phi_{\mathrm{C}}({\bvec{x}}) &= \frac{\omega_\mathrm{L} r^2}{2cR(\zeta)} + \Psi(\zeta),\\
        a_\mathrm{FF}({\bvec{x}}) &= a_0\left[1+\left(\frac{r}{w_\mathrm{FF}}\right)^2\right]\mathrm{exp}\left( -\frac{r^2}{w_\mathrm{FF}^2}\right), \label{eq:aFF} \\
        \phi_{\mathrm{FF}}({\bvec{x}}) &= 0, \label{eq:phiFF} 
\end{align}
where $\zeta = z - L_\mathrm{int}/2$ is the longitudinal position of the electron beam with respect to the center of the interaction region. The expressions for the conventional pulse correspond to a laser pulse focused by an ideal lens in the Gaussian optics approximation with a focal plane at $z = L_\mathrm{int}/2$, spot size 
$w = w_\mathrm{C}[1+(\zeta/Z_\mathrm{R})^2]^{1/2}$, radius of curvature $R(\zeta) = \zeta[1+Z_\mathrm{R}^2/\zeta^2]$, and Gouy phase $\Psi(\zeta)=\mathrm{arctan}(\zeta/Z_\mathrm{R})$. The focal plane was placed in the middle of the interaction region to (1) ensure the greatest amplitude uniformity, (2) provide the greatest average amplitude, and (3) allow the beam to radiate at all resonant wavelengths twice along its path (see Eq. \eqref{eq:Con_Res} and Fig. \ref{fig:2}b). The spatial variation in the phase $\phi_\mathrm{C}$ had almost no effect on the amplification for the simulated spot sizes.

{The expressions for the flying-focus pulse correspond to the ideal flying focus, which creates a moving focal point by focusing a laser pulse through a lens with a time-dependent focal length \cite{simpson2022spatiotemporal,ramsey2023exact}. The transverse profile of an ideal flying-focus pulse can be written as a superposition of any complete set of transverse modes\cite{ramsey2022nonlinear,formanek2022radiation,ramsey2023exact}. Here, the amplitude and phase are chosen to produce a flattened Gaussian beam (FGB) of order $N=1$ \cite{gori1994flattened}, which is a linear combination of the zeroth- and first-order radial Laguerre--Gaussian modes. The FGB profile was chosen to improve the transverse uniformity of the flying-focus undulator and weaken the transverse ponderomotive force on the electron beam. This could have also been achieved by using orthogonally polarized Laguerre-Gaussian modes with different orbital angular momentum values \cite{ramsey2020vacuum,ramsey2022nonlinear,formanek2023charged}.} 

{In general, the amplitude and phase of a flying-focus pulse are functions of $r$ and the moving coordinate $z-\varv_\mathrm{f} t$\cite{ramsey2023exact,formanek2022radiation}, which captures the effects of bandwidth on the LDFEL amplification process [Eqs. \eqref{eq:Ham} - \eqref{eq:parX}].} Because the highly relativistic electron beam is colocated with the moving focus and much shorter than the effective Rayleigh range $2\pi w_\mathrm{FF}^2/ \lambda_\mathrm{L}$ \cite{ramsey2023exact}, the $z$ and $t$-dependence in the amplitude and phase of the FGB is $\mathcal{O}(\lambda_\mathrm{L} L_\mathrm{b} /4\pi w_{\mathrm{FF}}^2) \sim \mathcal{O}(10^{-2})$ and is therefore neglected in Eqs. \eqref{eq:aFF} and \eqref{eq:phiFF}. In addition, the phase has an extrema at $r=0$ in the moving focal plane. Thus, the associated term in Eq. \eqref{eq:pondPhaseEvolve}, $d\phi_\mathrm{FF}/dz$, is negligible. The results of simulations with either the full expression for $\phi_\mathrm{FF}$ or $\phi_\mathrm{FF} = 0$ were identical. 

Wave propagation simulations of a conventional laser pulse with an $N =1$ FGB profile (not presented) showed a substantial increase in the amplitude nonuniformity across the interaction length compared to a pure Gaussian profile. Interference between the modes of the FGB resulted in two on-axis peaks located symmetrically about the focus ($z = L_\mathrm{int}/2 \pm 0.7 Z_\mathrm{R}$) with an intensity 1.3$\times$ larger than the intensity at focus. The interference also caused the flattopped transverse profile to rapidly degrade away from the focal plane. Due to the  exacerbated amplitude variation and degradation in the flattop profile, the conventional FGB was observed to produce a lower x-ray power than a conventional pulse with a simple Gaussian profile but the same energy. The interference between modes also occurs for an FGB flying focus, but the location of the intensity peaks and degraded flattop profile are located relative to the moving focal plane. As a result, an electron beam colocated and cotraveling with the moving focus does not enter the regions of space where these effects are prominent.

The energies of a conventional and flying-focus pulse are given by 
\begin{align}
   U_\mathrm{C} &= \frac{ mc^3}{16r_e}(a_0 k_\mathrm{L} w_\mathrm{C})^2 T, \\
  U_\mathrm{FF} &= \frac{\alpha mc^2}{8r_e}(a_0 k_\mathrm{L} w_\mathrm{FF})^2 L_\mathrm{f}
\end{align}
where $w_\mathrm{C}$ and $w_\mathrm{FF}$ are the $1/\mathrm{e}$ radii of the electric fields at focus for a Gaussian transverse profile, $T$ is the pulse duration, and $L_\mathrm{f} = cT/2$ when the focal velocity $\varv_\mathrm{f} = -\varv_\mathrm{ph} = c$ \cite{palastro2018ionization, ramsey2023exact,formanek2022radiation}. The pulse duration necessary to sustain the undulator over the interaction length $L_\mathrm{int}$ is $T=2L_\mathrm{int}/c$. For the simulated interaction, $L_\mathrm{int}=1.25$ cm, yielding $T=83$ ps as displayed in Table \ref{tab:SimParams}. A pure Gaussian transverse profile has an $\alpha = 1$, while an $N = 1$ FGB has $\alpha = 5/2$ [Eq. \eqref{eq:aFF}]. The $1/\mathrm{e}$ radius of the electric field for the full $N=1$ FGB profile at focus is 1.5$w_\mathrm{FF}$ or $56$ $\mu$m.

{As an alternative to the ideal flying focus used here, one could also consider using the chromatic flying focus \cite{froula2018spatiotemporal}. The chromatic flying focus creates a moving focal point by focusing a chirped laser pulse with a chromatic lens. In this case, the phase of the flying focus pulse is given by
\begin{equation}
\phi_{\mathrm{FF}} = \pm\frac{ \Delta \omega (z-L_\mathrm{f}/2)^2}{4cL_\mathrm{f}}
\end{equation}
where the $\pm$ refers to the sign of the chirp and $\Delta \omega$ is the bandwidth. The quadratic $z$-dependence of the phase corresponds to a frequency that varies linearly across the focal region $L_\mathrm{f}$. The resulting variation in the undulator period and detuning can be mitigated by modifying the focal geometry at the cost of more laser-pulse energy (see Discussion).}

%As the electron beam transverses the focal range, $L_\mathrm{f}$, the laser frequency it experiences changes linearly with distance.  As a result, the bandwidth of the chromatic flying focus causes frequency detuning. An LDFEL based on this flying focus implementation and the effects of the chirp are discussed in the fourth paragraph of the Discussion section. 

%As the electron beam transverses the focal range, the laser frequency shifts linearly. 

{In the simulations, the temporal profiles of both the flying focus and conventional pulse were assumed to be flat in the plane of the final focusing optics. This isolates the effects of amplitude nonuniformity introduced by diffraction from the effects of amplitude nonuniformity in the temporal profile. A varying temporal profile would add an additional source of detuning: $\Delta \lambda_\mathrm{X} {\sim} \int d\bvec{s} \cdot \nabla a^2_0\mathcal{T}^2(\bvec{x},t)$, where $\bvec{s}$ is the path of an electron through the undulator and $0\leq \mathcal{T}\leq 1$ is the temporal profile of the vector potential in the plane of the final focusing optic. Amplification requires a detuning $\Delta\lambda_\mathrm{X}/\lambda_{\mathrm{X}0}\lesssim 2\rho$, which can be reexpressed as a condition on the allowable variation in the temporal profile \cite{steiniger2014optical}
\begin{equation}
    \Delta \mathcal{T}^2 < 4\rho\frac{1+a_0^2/2}{a_0^2}.
\end{equation}
For the parameters in Table \ref{tab:SimParams}, $\Delta \mathcal{T}^2<1.1\%$ within the flat region of the pulse for both the conventional and flying focus pulse. However, when using a conventional pulse the temporal nonuniformity adds to the spatial nonuniformity, whereas the flying focus eliminates the spatial nonuniformity. In addition, smaller amplitudes relax the requirement on the temporal uniformity at the cost of longer saturation lengths, which favors the use of a flying-focus undulator (see Fig. \ref{fig:3}).} 

\section*{Data availability}
Data underlying the results presented in this paper are not publicly available at this time, but may be obtained from the corresponding author upon reasonable request.

\section*{Code availability}
Code used to generate data is not publicly available at this time, but can be obtained from the corresponding author upon reasonable request.

%\bibliography{LDFEL}
%\input{LDFEL.bbl} 

\section*{Acknowledgments}
The authors extend their gratitude to A. Di Piazza, K. G. Miller, W. Scullin, K. Weichman, A. L. Elliott, J. Maxson, G. Bruhaug, and G. W. Collins for their insight and engaging discussions.

This report was prepared as an account of work sponsored by an agency of the US Government. Neither the US Government nor any agency thereof, nor any of their employees, makes any warranty, express or implied, or assumes any legal liability or responsibility for the accuracy, completeness, or usefulness of any information, apparatus, product, or process disclosed, or represents that its use would not infringe privately owned rights. Reference herein to any specific commercial product, process, or service by trade name, trademark, manufacturer, or otherwise does not necessarily constitute or imply its endorsement, recommendation, or favoring by the US Government or any agency thereof. The views and opinions of authors expressed herein do not necessarily state or reflect those of the US Government or any agency thereof. 

This material is based upon work supported by the Department of Energy [National Nuclear Security Administration] University of Rochester ``National Inertial Confinement Fusion Program'' under Award Number DE-NA0004144 and Department of Energy Office of Science under Award Number DE-SC0021057. The work of M.F. is supported by the European Union’s Horizon Europe research and innovation program under the Marie Sklodowska-Curie Grant Agreement No. 101105246STEFF.

\section*{Author contributions statement}
D.R., D.H.F., and J.P.P. conceived the flying-focus undulator configuration. D.R. developed and performed simulations. D.R. and J.P.P. developed the theory and analyzed simulation data. B.M., T.T.S., M.F., L.S.M, and J.V. provided theoretical and numerical expertise. T.T.S., D.H.F., and J.P.P. provided experimental insights. D.R. and J.P.P. wrote the manuscript. All authors reviewed and edited the manuscript. 

\section*{Ethics declarations}
\subsection*{Competing interest}
The authors declare no competing interests.

\end{document}